\documentclass[a4paper,11pt]{article}
\usepackage{jheppub} % for details on the use of the package, please see the JINST-author-manual
\usepackage{lineno}

\arxivnumber{2308.13518}

\title{Holographic Euclidean thermal correlator}

\author[a,b]{Song He}
\author[a]{and Yi Li}
\affiliation[a]{
 Center for Theoretical Physics and College of Physics, Jilin University, \\Changchun 130012, People's Republic of China
}
\affiliation[b]{
 Max Planck Institute for Gravitational Physics (Albert Einstein Institute), \\Am M\"uhlenberg 1, 14476 Golm, Germany
}

\emailAdd{hesong@jlu.edu.cn, liyi@fudan.edu.cn}

\abstract{In this paper, we compute holographic Euclidean thermal correlators of the stress tensor and $U(1)$ current from the AdS planar black hole. To this end, we set up perturbative boundary value problems for Einstein's gravity and Maxwell theory in the spirit of Gubser-Klebanov-Polyakov-Witten, with appropriate gauge fixing and regularity boundary conditions at the horizon of the black hole. The linearized Einstein equation and Maxwell equation in the black hole background are related to the Heun equation of degenerate local monodromy. Leveraging the connection relation of local solutions of the Heun equation, we partly solve the boundary value problem and obtain exact two-point thermal correlators for $U(1)$ current and stress tensor in the scalar and shear channels.}

\begin{document}
\maketitle
\flushbottom

\section{Introduction}
\label{sec:introduction}
As an embodiment of the holographic principle \cite{tHooft:1993dmi, Susskind:1994vu}, the Anti-de Sitter gravity/conformal field theory (AdS/CFT) correspondence \cite{Maldacena:1997re, Gubser:1998bc, Witten:1998qj} establishes a connection between a quantum gravity theory in AdS space and a conformal field theory on the boundary. This equivalence is encapsulated in the Gubser-Klebanov-Polyakov-Witten (GKPW) relation, where the partition function of the conformal field theory with operator sources equals the gravity partition function with prescribed boundary conditions
\begin{align}
    \langle e^{\int \phi_0 O} \rangle_{CFT} = Z_{\mathrm{G}}[\phi_0]
\end{align}
In the most helpful limit to exploit this correspondence, the classical gravity on-shell action becomes the generating functional of connected correlators of the strongly-coupled CFT
\begin{align}
    I_{CFT}[\phi_0] = I_{\mathrm{G,on-shell}}[\phi_0]
\end{align}
Correlators are computed by functional differentiation of the generating functional, which amounts to solving the perturbative boundary value problem for the bulk fields' equation of motion. This involves varying the boundary value of the bulk fields and solving for the corresponding variation of the on-shell configuration in the bulk. The near-boundary behavior is well-established, allowing the extraction of holographic correlators \cite{Fefferman:1985cfm, deHaro:2000vlm, Bianchi:2001kw, Fefferman:2007rka}. However, solving the global boundary value problem is generally intricate, exemplified in cases like pure gravity \cite{Anderson:2004yi}.

Although the prescription is clear, explicit computation of holographic Euclidean correlators in the GKPW approach has been limited to pure AdS space and its quotient spaces, such as thermal AdS where the method of images can be applied (e.g., see \cite{Alday:2020eua} for thermal bootstrap emphasis). In our prior work \cite{He:2023hoj}, we computed holographic torus correlators of the stress tensor. This study focuses on Euclidean thermal two-point correlators of the stress tensor and $U(1)$ current in four-dimensional CFTs. Beyond the Hawking-Page transition \cite{Hawking:1982dh}, the thermal state holographically corresponds to a five-dimensional Euclidean AdS planar black hole \cite{Witten:1998zw}. Correlators are derived by solving perturbative boundary value problems in Einstein's gravity and Maxwell theory for the $U(1)$ gauge field in the black hole background. Two important steps are involved. The first is to appropriately fix the gauge and impose regularity boundary conditions at the horizon, ensuring a unique solution. The second step identifies the equations of motion as the Heun equation \cite{Heun:1888zt}, and solves the boundary value problems with the connection relation of local solutions. The general connection relation was established in \cite{Bonelli:2022ten}, and it has been applied to exact thermal correlators in Minkowski signature \cite{Dodelson:2022yvn, Bhatta:2022wga}, and employed in various black hole perturbation problems \cite{Aminov:2020yma,Bonelli:2021uvf,Consoli:2022eey,Giusto:2023awo,Aminov:2023jve,Lei:2023mqx}. In our case, the Heun equations feature degenerate local monodromy, with characteristic exponents differing by an integer. We compute the connection relation by taking a limit of the generic case. Ultimately, we obtain exact two-point correlators for the $U(1)$ current and stress tensor in the scalar and shear channels (as defined in \cite{Kovtun:2005ev}).

Thermal two-point correlators, also known as thermal spectral functions, have many important applications and have been studied in \cite{Kovtun:2005ev} using gauge invariants in each channel. In the final discussion section, we comment on our approach to holographic computation and relevant applications of thermal two-point correlators.

\section{Holographic setup}
\label{sec:holographic setup}
We start by reviewing the basics of holographic computation independent of the bulk background geometry. For Einstein's gravity, it's customary to work in the Fefferman-Graham gauge \cite{Fefferman:1985cfm, Graham:1991jqw} near the conformal boundary
\begin{align} \label{gravity FG gauge}
 ds^2 = \frac{dr^2}{r^2} + \frac{1}{r^2} \mathbf{g}_{ij}(r,x) dx^i dx^j
\end{align}
and in dimension four, we have the series expansion
\begin{align} \label{gravity FG series}
    \mathbf{g}_{ij} = \mathbf{g}^{(0)}_{ij} + r^2 \mathbf{g}^{(2)}_{ij} + r^4 \mathbf{g}^{(4)}_{ij} + r^4 \log r \mathbf{h}^{(4)}_{ij} + \ldots
\end{align}
The background metric of the holographic field theory $\gamma_{ij}$ corresponds to $\mathbf{g}^{(0)}_{ij}$, and the one-point correlator of the stress tensor\footnote{Our convention for the stress tensor is, for classical field theories, $\delta_\gamma S = \frac{1}{2}\int dV T_{ij} \delta \gamma^{ij}$. For quantum theories $\langle T_{ij}\rangle = 2 \frac{\delta I}{\delta \gamma^{ij}}$. The integration measure is assumed to be included in the definition of the functional derivative.}, with appropriate renormalization, is given by \cite{Emparan:1999pm,Kraus:1999di,deHaro:2000vlm}
\begin{align} \label{stress tensor 1pt correlator}
    \langle T_{ij}\rangle = \frac{4}{16\pi G}\big[\mathbf{g}^{(4)}_{ij} - \frac{1}{8}\mathbf{g}^{(0)}_{ij}(\mathbf{P}^{(0)2}-\mathbf{P}^{(0)}_{kl}\mathbf{P}^{(0)kl}) - \frac{1}{2}\mathbf{P}^{(0)}_{ik}\mathbf{P}^{(0)k}_j + \frac{1}{4}\mathbf{P}^{(0)}\mathbf{P}^{(0)}_{ij}\big]
\end{align}
where $\mathbf{P}^{(0)}_{ij}$ is the Schouten tensor of $\mathbf{g}^{(0)}_{ij}$. In our case, the holographic field theory lives on a flat background, so the terms of Schouten tensor do not contribute. We have 
\begin{align} \label{stress tensor 1pt correlator simplified}
    \langle T_{ij}\rangle = \frac{1}{4\pi G} \mathbf{g}^{(4)}_{ij}
\end{align}
The Einstein equation near the conformal boundary determines the series (\ref{gravity FG series}) in terms of $\mathbf{g}^{(0)}_{ij}$ and $\mathbf{g}^{(4)}_{ij}$ or equivalently the one point correlator $\langle T_{ij}\rangle$, and imposes holographic Ward identities of conservation and the Weyl anomaly on $\langle T_{ij}\rangle$. Near boundary solutions of the Einstein equation are in one-to-one correspondence to the pair $(\gamma_{ij},\langle T_{ij}\rangle)$. The global geometry of the bulk spacetime fully determines the one-point correlator as a functional of the boundary metric, from which we can compute multi-point correlators by functional differentiation. In particular for two-point correlator we have
\begin{align} \label{T 1pt to 2pt}
    \langle T_{ij}(x) T_{kl}(y)\rangle = -2 \frac{\delta \langle T_{ij}(x) \rangle}{\delta \gamma^{kl}(y)}
\end{align}

Similarly, for the $U(1)$ gauge field, we can put it in the radial gauge near the conformal boundary using the Fefferman-Graham coordinates of the bulk metric
\begin{align}
 A = \mathbf{A}_i(r,x) dx^i
\end{align}
with the series expansion
\begin{align} \label{gauge field FG series}
    \mathbf{A}_i = \mathbf{A}^{(0)}_i + r^2 \mathbf{A}^{(2)}_i + r^2 \log r \mathbf{B}^{(2)}_i + \ldots
\end{align}
The one-point correlator with appropriate renormalization is given by
\begin{align} \label{J renormalized 1pt correlator}
    \langle J_i \rangle = - 2 \mathbf{A}^{(2)}_i
\end{align}
The Maxwell equation near the conformal boundary determines the series (\ref{gauge field FG series}) in terms of $\mathbf{A}^{(0)}_i$ and $\mathbf{A}^{(2)}_i$ or equivalently the one point correlator $\langle J_i\rangle$, and imposes the holographic Ward identity of conservation on $\langle J_i\rangle$. The global geometry of the bulk spacetime fully determines $\langle J_i \rangle$. Multi-point correlators are computed by taking functional derivative with respect to the source $\mathcal{A}$, for two point correlator we have
\begin{align} \label{J 1pt to 2pt}
    \langle J_i(x) J_j(y) \rangle = \frac{\delta \langle J_i(x) \rangle}{\delta \mathcal{A}^j(y)}
\end{align}

Now we specialize in the holographic correlators from the five-dimensional AdS planar black hole. The black hole is a solid cylinder $\mathbb{B}^2 \times \mathbb{R}^3$ with the metric
\begin{align}  \label{AdS planar black hole metric}
    ds^2 = \frac{1}{\rho^2}[(1-\frac{\rho^4}{\rho_0^4})^{-1}d\rho^2 + (1-\frac{\rho^4}{\rho_0^4})dt^2 + d\Vec{x}^2]
\end{align}
The period of Euclidean time $t$, namely the inverse temperature, is $\beta=\pi\rho_0$. The conformal boundary is at $\rho=0$, and the horizon is at $\rho=\rho_0$, being the $\mathbb{R}^3$ axis of the cylinder. The standard Fefferman-Graham radial coordinate $r$ is related to $\rho$ by
\begin{align}
    \rho = \frac{r}{\sqrt{1+\frac{r^4}{4\rho_0^4}}}
\end{align}
For simplicity, we set $\rho_0=1$ in the metric, effectively working in the unit of $\rho_0$, and we will recover $\rho_0$ dependence when final results are obtained. As a convention, we label bulk spacetime coordinate indices by Greek alphabets $\mu,\nu,\rho,\ldots$, the boundary spacetime coordinate indices by Roman alphabets $i,j,k,\ldots$ and the boundary space indices by $a,b,c,\ldots$.

\section{$U(1)$ current}
\label{sec:U(1) current}
Now, we work on the boundary value problem of the $U(1)$ gauge field, beginning with gauge-fixing. We can put it in the radial gauge $A_\rho=0$ in the region $0\leq\rho<1$ (excluding the horizon) by a $U(1)$ gauge transformation. For a global solution, its restriction to the region $0\leq\rho<1$ must have a regular limit going to the horizon $\rho=1$. Therefore, we formulate the boundary value problem in the radial gauge, with the boundary condition that the solution has a regular limit as $\rho\to 1$ after a gauge transformation. To work out the explicit form of the boundary condition, we introduce the ``cylindrical radial coordinate" $\mathfrak{s}$
\begin{align}
    \cosh 2\mathfrak{s} = \frac{1}{\rho^2}
\end{align}
Near the horizon $\rho\to 1$ or $\mathfrak{s}\to 0$, the metric takes the form of Euclidean metric in ``cylindrical coordinates"
\begin{align}
    ds^2 \sim d\mathfrak{s}^2 + \mathfrak{s}^2 d(2t)^2 + d\vec{x}^2
\end{align}
and the horizon is properly covered by the ``Cartesian coordinates"
\begin{align} \label{horizon Cartesian}
    X &=\mathfrak{s}\cos 2t \notag\\
    Y &=\mathfrak{s}\sin 2t \notag\\
    \vec{x} &= \vec{x}
\end{align}
The gauge field is regular at the horizon if and only if its components in a coordinate chart that properly covers the horizon, for example the ``Cartesian coordinates", are regular. That is, we have
\begin{align}
    A = \mathbf{A}_i dx^i
\end{align}
and there exists a $U(1)$ gauge transformation $\Lambda$, such that
\begin{align}
\lim_{\mathfrak{s}\to 0 } A + d\Lambda = A^*_X(\vec{x}) dX + A^*_Y(\vec{x}) dY + A^*_a(\vec{x}) dx^a 
\end{align}
The components on the right-hand side can only depend on $\vec{x}$ because the $t$-circle shrinks to a point as $\mathfrak{s}\to 0$. We find
\begin{align}
 &\lim_{\mathfrak{s}\to 0} \partial_\mathfrak{s} \Lambda = A^*_X(\vec{x}) \cos 2t + A^*_Y(\vec{x}) \sin 2t \label{gauge field regularity eqn 1}\\
 &\lim_{\mathfrak{s}\to 0} \frac{\mathbf{A}_t + \partial_t \Lambda}{\mathfrak{s}} = -2A^*_X(\vec{x}) \sin 2t + 2A^*_Y(\vec{x}) \cos 2t \label{gauge field regularity eqn 2}\\
 &\lim_{\mathfrak{s}\to 0} \mathbf{A}_a + \partial_a \Lambda = A^*_a(\vec{x}) \label{gauge field regularity eqn 3}
\end{align}
From (\ref{gauge field regularity eqn 1}) we see $\Lambda$ can be approximated as a linear function of $\mathfrak{s}$ as $\mathfrak{s}\to 0$ (or $\rho\to 1$), then from (\ref{gauge field regularity eqn 3}) we know
\begin{align} \label{gauge field regularity condition 1}
    \lim_{\rho\to 1} \mathbf{A}_a \; \mathrm{exists}
\end{align}
In addition, by integrating (\ref{gauge field regularity eqn 2}) over $t$ we find
\begin{align} \label{gauge field regularity condition 2}
    \int_0^\pi dt \mathbf{A}_t|_{\rho=1} = 0
\end{align}
This gauge fixing and regularity boundary conditions at the horizon, together with the boundary value
\begin{align}
    {\mathbf{A}_i}|_{\rho=0} &= \mathcal{A}_i
\end{align}
as a turned-on source on the CFT side, determine a unique solution to the Maxwell equation as we will see.

We utilize the translational symmetry in $t,\vec{x}$ direction and work with Fourier modes $\tilde{\mathbf{A}}_i$ with Matsubara frequency $\omega=2m,m\in \mathbb{Z}$ and spatial momentum $\vec{p}$. For simplicity, we also rotate the spatial momentum to the $x^1$ direction. The Maxwell equation
\begin{align} \label{Maxwell eqn}
    d\ast F = 0
\end{align}
then decouples to the transverse channel for $\tilde{\mathbf{A}}_2,\tilde{\mathbf{A}}_3$ and the longitudinal channel for $\tilde{\mathbf{A}}_t,\tilde{\mathbf{A}}_1$. For the transverse component $\tilde{\mathbf{A}}_2$ (and the same for $\tilde{\mathbf{A}}_3$) we have
\begin{align}
    (\partial_z^2 - \frac{2z}{1-z^2}\partial_z - \frac{4m^2+p^2(1-z^2)}{4z(1-z^2)^2}) \tilde{\mathbf{A}}_2 = 0
\end{align}
where we used the convenient coordinate $z=\rho^2$. This is an ordinary differential equation with four regular singularities $z=0,1,-1,\infty$. By the substitution $\tilde{\mathbf{A}}_2(z) = (1-z^2)^{-\frac{1}{2}} w(z)$, we get a Heun equation in the normal form for $w(z)$
\begin{align}
    (\partial_z^2 + \frac{\frac{1}{4}-(\frac{1}{2})^2}{z^2} + \frac{\frac{1}{4}-(\frac{m}{2})^2}{(z-1)^2} + \frac{\frac{1}{4}-(\frac{m}{2}i)^2}{(z+1)^2} + \frac{p^2+4m^2-2}{8z(z-1)} -\frac{p^2+4m^2+2}{8z(z+1)})w(z)=0
\end{align}
with the Heun equation parameters
\begin{align}
    t=-1,a_0=\frac{1}{2},a_1=\frac{|m|}{2},a_t=\frac{m}{2}i,a_\infty=\frac{1}{2},u=-
\frac{p^2+4m^2+2}{8}
\end{align}
We refer the readers to Appendix \ref{Fuchsian ODE, Heun and connection problem} for a brief review of Fuchsian differential equations, the Heun equation, its connection problem, and notational conventions.
By the boundary condition (\ref{gauge field regularity condition 1}), $\tilde{\mathbf{A}}_2$ is regular at $z=1$, so it must be proportional to the solution of exponent $\frac{|m|}{2}$ at $z=1$. The constant of proportionality is determined by the boundary condition $\tilde{\mathbf{A}}_2|_{z=0}=\tilde{\mathcal{A}}_2$ and the connection relation (\ref{connection relation a0=1/2}). We find
\begin{align}
    &\tilde{\mathbf{A}}_2(\omega=2m,p,z) = \tilde{\mathcal{A}}_2(\omega,p) (1-z^2)^{-\frac{1}{2}}\big[ w^{(0)}_{-} + \frac{p^2+4m^2}{4}
    (-2\psi(1)-1 \nonumber\\
    &+\frac{1}{2} \sum_{\theta,\sigma=\pm}\psi(\theta\frac{m}{2}+\sigma a) - \frac{1}{2} \partial_{a_0}^2 F|_{a_0=\frac{1}{2}} - \frac{2}{p^2+4m^2}(1+2\partial_t\partial_{a_0}F|_{a_0=\frac{1}{2},t=-1})) w^{(0)}_{+} \big]
\end{align}

For the longitudinal components $\tilde{\mathbf{A}}_t,\tilde{\mathbf{A}}_1$, we have
\begin{align}
    &\partial_z^2 \tilde{\mathbf{A}}_t - \frac{p^2}{4z(1-z^2)} \tilde{\mathbf{A}}_t + \frac{2mp}{4z(1-z^2)} \tilde{\mathbf{A}}_1 = 0 \label{gauge field longitudinal eq1}\\
    &\partial_z^2 \tilde{\mathbf{A}}_1 - \frac{2z}{1-z^2}\partial_z \tilde{\mathbf{A}}_1 - \frac{4m^2}{4z(1-z^2)^2}\tilde{\mathbf{A}}_1 + \frac{2mp}{4z(1-z^2)^2} \tilde{\mathbf{A}}_t = 0 \label{gauge field longitudinal eq2}\\
    &\frac{2m}{1-z^2}\partial_z \tilde{\mathbf{A}}_t + p\partial_z \tilde{\mathbf{A}}_1 = 0 \label{gauge field longitudinal eq3}
\end{align}
Plugging (\ref{gauge field longitudinal eq3}) into $\partial_z\big(z(1-z^2)$(\ref{gauge field longitudinal eq1})$\big)$, we obtain
\begin{align} \label{gauge field At eqn}
    (\partial_z^2 + \frac{1-3z^2}{z(1-z^2)} \partial_z - \frac{p^2(1-z^2)+4m^2}{4z(1-z^2)^2})\partial_z \tilde{\mathbf{A}}_t = 0
\end{align}
When $m\neq 0$, the solution to this third-order differential equation is  determined by the three boundary conditions
\begin{align}
    &\tilde{\mathbf{A}}_1|_{z=1}\; \mathrm{regular} \nonumber\\
    &\tilde{\mathbf{A}}_1|_{z=0} = \tilde{\mathcal{A}}_1,\; \tilde{\mathbf{A}}_t|_{z=0} = \tilde{\mathcal{A}}_t
\end{align}
By the substitution $\partial_z \tilde{\mathbf{A}}_t = z^{-\frac{1}{2}}(1-z^2)^{-\frac{1}{2}} w(z)$, (\ref{gauge field At eqn}) can be transformed to the normal Heun equation
\begin{align}
    \big(\partial_z^2 + \frac{\frac{1}{4}-0^2}{z^2} + \frac{\frac{1}{4}-(\frac{m}{2})^2}{(z-1)^2} + \frac{\frac{1}{4}-(\frac{m}{2}i)^2}{(z+1)^2} + \frac{p^2 + 4m^2 -6}{8z(z-1)} - \frac{p^2 + 4m^2 +6}{8z(z+1)} \big)w(z) = 0
\end{align}
with
\begin{align}
    t=-1,a_0=0, a_1=\frac{|m|}{2}, a_t=\frac{m}{2}i, a_\infty=1, u=-\frac{p^2+4m^2+6}{8}
\end{align}
By (\ref{gauge field longitudinal eq3}) the solution must be proportional to $w^{(1)}_{+}$ for $\tilde{\mathbf{A}}_1$ to be regular at $z=1$. The constant of proportionality can be further determined by using the connection relation (\ref{connection relation a0=1/2}) and evaluating (\ref{gauge field longitudinal eq1}) at $z=0$. We find
\begin{align}
    z^\frac{1}{2}\sqrt{1-z^2}\partial_z \tilde{\mathbf{A}}_t = &\frac{2mp\tilde{\mathcal{A}}_1-p^2\tilde{\mathcal{A}}_t}{4} \big[ -w^{(0)}_{-}(z) \nonumber\\
    &+ (2\psi(1)-\frac{1}{2} \sum_{\theta,\sigma=\pm}\psi(\frac{1}{2}+\theta\frac{m}{2}+\sigma a) +\frac{1}{2}\partial_{a_0}^2F)w^{(0)}_{+} \big]
\end{align}
Then we integrate to obtain $\tilde{\mathbf{A}}_t$ with the constant of integration given by the boundary value $\tilde{\mathcal{A}}_t$
\begin{align}
    \tilde{\mathbf{A}}_t = &\tilde{\mathcal{A}}_t+\frac{2mp\tilde{\mathcal{A}}_1-p^2\tilde{\mathcal{A}}_t}{4} \big[ -(z\log z+\ldots) \nonumber\\
    &+ (2\psi(1)+1-\frac{1}{2} \sum_{\theta,\sigma=\pm}\psi(\frac{1}{2}+\theta\frac{m}{2}+\sigma a)+\frac{1}{2}\partial_{a_0}^2F)(z+\ldots)) \big] 
\end{align}
We get $\tilde{\mathbf{A}}_1$ by plugging $\tilde{\mathbf{A}}_t$ back to (\ref{gauge field longitudinal eq3})
\begin{align}
    \tilde{\mathbf{A}}_1 = &\tilde{\mathcal{A}}_1 (1+\ldots) + \frac{2m(p\tilde{\mathcal{A}}_t-2m\tilde{\mathcal{A}}_1)}{4} \nonumber\\
    &\times\big(2\psi(1)+1 -\frac{1}{2} \sum_{\theta,\sigma=\pm}\psi(\frac{1}{2}+\theta\frac{m}{2}+\sigma a)+\frac{1}{2}\partial_{a_0}^2F\big)(z+\ldots)
\end{align}
When $m=0$, we get $\tilde{\mathbf{A}}_1=\tilde{\mathcal{A}}_1$ from (\ref{gauge field longitudinal eq3}). We still solve for $z^\frac{1}{2}\sqrt{1-z^2}\partial_z \tilde{\mathbf{A}}_t$ from (\ref{gauge field At eqn}), which is a linear combination of $w^{(1)}_{+}=\sqrt{1-z}(1+\ldots)$ and $w^{(1)}_{-}=\sqrt{1-z}(\log(1-z)+\ldots)$. Then we plug it into (\ref{gauge field longitudinal eq1}) and evaluate at $z=1$. We have ${\tilde{\mathbf{A}}_t}(m=0)|_{z=1} = 0$ from the boundary condition (\ref{gauge field regularity condition 2}), and we find $z^\frac{1}{2}\sqrt{1-z^2}\partial_z \tilde{\mathbf{A}}_t$ must be proportional to $w^{(1)}_{+}$, the same as the previous case when $m\neq 0$. So, we can carry over the results for $m\neq0$ and set $m=0$ in the expression. 

To obtain the holographic correlators, we recover the dependence on $\rho_0$ or the inverse temperature $\beta=\pi\rho_0$, and read off $\mathbf{A}^{(2)}_i$ from the bulk gauge field $\mathbf{A}_i$ (in our case the coefficient of $z^1$)
\begin{align}
    \tilde{\mathbf{A}}_2^{(2)}(\omega=\frac{2m}{\rho_0},p) &= -\frac{p^2+\omega^2}{4}\tilde{\mathcal{A}}_2(\omega,p) \mathcal{C}_1(\omega=\frac{2m}{\rho_0},p) \nonumber\\
    \tilde{\mathbf{A}}_t^{(2)}(\omega=\frac{2m}{\rho_0},p) &= \frac{\omega p \tilde{\mathcal{A}}_1(\omega,p)-p^2\tilde{\mathcal{A}}_t(\omega,p)}{4} \mathcal{C}_2(\omega=\frac{2m}{\rho_0},p) \nonumber\\
    \tilde{\mathbf{A}}_1^{(2)}(\omega=\frac{2m}{\rho_0},p) &= \frac{\omega p \tilde{\mathcal{A}}_t(\omega,p) - \omega^2 \tilde{\mathcal{A}}_1(\omega,p)}{4} \mathcal{C}_2(\omega=\frac{2m}{\rho_0},p)
\end{align}
where
\begin{align}
    &\mathcal{C}_1(\omega=\frac{2m}{\rho_0},p)=(2\psi(1)+1-\frac{1}{2} \sum_{\theta,\sigma=\pm}\psi(\theta\frac{m}{2}+\sigma a) \nonumber\\
    &+\frac{1}{2}\partial_{a_0}^2F+\frac{2}{\rho_0^2p^2+4m^2}(1+\partial_t\partial_{a_0}F))\big|_{t=-1,a_0=\frac{1}{2},a_1=\frac{|m|}{2},a_t=\frac{m}{2}i,a_\infty=\frac{1}{2},u=-\frac{\rho_0^2p^2+4m^2+2}{8}} \nonumber\\
    &\mathcal{C}_2(\omega=\frac{2m}{\rho_0},p)=(2\psi(1)+1-\frac{1}{2} \sum_{\theta,\sigma=\pm}\psi(\frac{1}{2}+\theta\frac{m}{2}+\sigma a)\nonumber\\
    &+\frac{1}{2}\partial_{a_0}^2F)\big|_{t=-1,a_0=0,a_1=\frac{|m|}{2},a_t=\frac{m}{2}i,a_\infty=1,u=-\frac{\rho_0^2p^2+4m^2+6}{8}}
\end{align}
We compute two-point correlators by the formula for renormalized one-point correlators (\ref{J renormalized 1pt correlator}). Rotating the spatial momentum to a general direction, we find
\begin{align}
    \langle \tilde{J}_t(\omega,p) \tilde{J}_t(-\omega,-p)\rangle &=\frac{p^2}{2}\mathcal{C}_2(\omega,p) \nonumber\\
    \langle \tilde{J}_t(\omega,p) \tilde{J}_b(-\omega,-p)\rangle &= -\frac{\omega}{2}\mathcal{C}_2(\omega,p)p_b \nonumber\\
    \langle \tilde{J}_a(\omega,p) \tilde{J}_b(-\omega,-p)\rangle &= \frac{p^2+\omega^2}{2}\mathcal{C}_1(\omega,p)(\delta_{ab}-\frac{p_ap_b}{p^2}) + \frac{\omega^2}{2}\mathcal{C}_2(\omega,p)\frac{p_ap_b}{p^2}
\end{align}

\section{Stress tensor}
\label{sec:stress tensor}
Gauge fixing and regularity boundary conditions at the horizon for Einstein's gravity follow the same line as the Maxwell theory. We can make the solid cylinder coordinates $\rho,t,\vec{x}$ the Fefferman-Graham coordinates of the perturbed bulk metric in the region $0\leq\rho<1$ by a diffeomorphism. Then, the boundary value problem is formulated in this gauge with the boundary condition that the metric has a regular limit as $\rho\to 1$ after a diffeomorphism. For a first-order perturbation of the bulk metric, we have
 \begin{align}
     \delta ds^2 =\delta \mathbf{g}_{ij} dx^i dx^j
 \end{align}
 And to the first order, the diffeomorphism is characterized by a vector $V$, then the regularity boundary condition at the horizon is the variation of the bulk metric
 \begin{align}
     \mathcal{L}_V (ds^2) + \delta ds^2
 \end{align}
has a regular limit as $\rho\to 1$ (or $\mathfrak{s}\to 0$), that is, its components in the "Cartesian coordinates" (\ref{horizon Cartesian}) are regular. We find
\begin{align}
    &\lim_{\mathfrak{s}\to 0}2\partial_\mathfrak{s} V^\mathfrak{s} = \cos^22t \delta g^*_{XX} + 2\cos2t\sin2t \delta g^*_{XY} + \sin^22t \delta g^*_{YY} \label{gravity regularity eqn 1}\\
    &\lim_{\mathfrak{s}\to 0} \frac{\partial_t V^\mathfrak{s} + \frac{\sinh^22\mathfrak{s}}{\cosh2\mathfrak{s}}\partial_\mathfrak{s} V^t}{2\mathfrak{s}} \nonumber\\
    &= -\cos2t\sin2t \delta g^*_{XX} + (\cos^22t-\sin^22t)\delta g^*_{XY} + \cos2t\sin2t \delta g^*_{YY} \label{gravity regularity eqn 2}\\
    &\lim_{\mathfrak{s}\to 0}\partial_a V^\mathfrak{s} + \cosh2\mathfrak{s}\partial_\mathfrak{s} V^x = \cos2t \delta g^*_{Xa} + \sin2t \delta g^*_{Ya} \label{gravity regularity eqn 3}\\
    &\lim_{\mathfrak{s}\to 0}\frac{\delta \mathbf{g}_{tt} + \frac{\sinh2\mathfrak{s}}{\cosh^22\mathfrak{s}}((3+4\cosh4\mathfrak{s})V^\mathfrak{s}+\sinh4\mathfrak{s}\partial_t V^t)}{4\mathfrak{s}^2} \nonumber\\
    &= \sin^22t \delta g^*_{XX} - 2\cos2t\sin2t \delta g^*_{XY} + \cos^22t \delta g^*_{YY} \label{gravity regularity eqn 4}\\
    &\lim_{\mathfrak{s}\to 0}\delta \mathbf{g}_{ta} + \frac{\sinh^22\mathfrak{s}}{\cosh2\mathfrak{s}}\partial_a V^t + \cosh2\mathfrak{s}\partial_t V^a = -2\mathfrak{s}\sin2t \delta g^*_{Xa} + 2\mathfrak{s}\cos2t \delta g^*_{Ya} \label{gravity regularity eqn 5}\\
    &\lim_{\mathfrak{s}\to 0}\delta \mathbf{g}_{ab} + \cosh2\mathfrak{s}(\partial_a V^b + \partial_b V^a) + 2\sinh2\mathfrak{s} V^\mathfrak{s} \delta_{ab} = \delta g^*_{ab} \label{gravity regularity eqn 6}
\end{align}
(\ref{gravity regularity eqn 1}) and (\ref{gravity regularity eqn 3}) show that $V^\mathfrak{s}$ and $V^a$ can be approximated by linear function in $\mathfrak{s}$ as $\mathfrak{s}\to 0$ (or $\rho\to 1$). Plugging into (\ref{gravity regularity eqn 6}), we see 
\begin{align} \label{gravity regularity condition 1}
   \lim_{\rho\to 1}\delta \mathbf{g}_{ab} \; \mathrm{exists}
\end{align}
By (\ref{gravity regularity eqn 2}) we know
\begin{align}
    V^t = O(\frac{1}{\mathfrak{s}})
\end{align}
as $\mathfrak{s}\to 0$. Then by integrating (\ref{gravity regularity eqn 5}) over $t$ we find 
\begin{align} \label{gravity regularity condition 2}
\int_0^\pi dt \delta \mathbf{g}_{ta}|_{\rho=1} = 0
\end{align}
 
Similar to the case of $U(1)$ gauge field, we work in Fourier modes and rotate the spatial momentum to the $x^1$ direction. And for simplicity, we use the variable $\mathbf{h}_{ij} = \rho^2\delta \mathbf{g}_{ij}$ which on the conformal boundary equals the variation of the CFT background metric
\begin{align}
    {\mathbf{h}_{ij}}|_{\rho=0} = \delta \gamma_{ij}
\end{align}
The linearized Einstein equation
\begin{align} \label{linearized Einstein eqn}
    \frac{1}{2}(\nabla^\lambda \nabla_\mu \delta g_{\lambda\nu} + \nabla^\lambda \nabla_\nu \delta g_{\lambda\mu} - \nabla^\lambda\nabla_\lambda \delta g_{\mu\nu} - \nabla_\mu \nabla_\nu \delta g^\lambda_\lambda) + 4 \delta g_{\mu\nu} = 0
\end{align}
decouples to the scalar channel of $\tilde{\mathbf{h}}_{23}$ and $\tilde{\mathbf{h}}_{22}-\tilde{\mathbf{h}}_{33}$, the shear channel of $\tilde{\mathbf{h}}_{t2},\tilde{\mathbf{h}}_{12}$ and $\tilde{\mathbf{h}}_{t3},\tilde{\mathbf{h}}_{13}$, and the sound channel of $\tilde{\mathbf{h}}_{tt},\tilde{\mathbf{h}}_{11},\tilde{\mathbf{h}}_{22}+\tilde{\mathbf{h}}_{33},\tilde{\mathbf{h}}_{t1}$.
In the scalar channel, we have
\begin{align}
    \partial_z^2 \tilde{\mathbf{h}}_{23} - \frac{1+z^2}{z(1-z^2)}\partial_z \tilde{\mathbf{h}}_{23} - \frac{p^2(1-z^2)+\omega^2}{4z(1-z^2)^2} \tilde{\mathbf{h}}_{23} =0
\end{align}
and in the shear channel, we have
\begin{align}
    &\partial_z^2 \tilde{\mathbf{h}}_{t2} - \frac{1}{z}\partial_z \tilde{\mathbf{h}}_{t2} - \frac{p^2}{4z(1-z^2)}\tilde{\mathbf{h}}_{t2} + \frac{2mp}{4z(1-z^2)} \tilde{\mathbf{h}}_{12}=0 \label{gravity shear eqn1}\\
    &\partial_z^2 \tilde{\mathbf{h}}_{12} - \frac{1+z^2}{z(1-z^2)}\partial_z \tilde{\mathbf{h}}_{12} - \frac{4m^2}{4z(1-z^2)^2} \tilde{\mathbf{h}}_{12} + \frac{2mp}{4z(1-z^2)^2} \tilde{\mathbf{h}}_{t2} = 0 \label{gravity shear eqn2}\\
    &\frac{2m}{1-z^2} \partial_z \tilde{\mathbf{h}}_{t2} + p \partial_z \tilde{\mathbf{h}}_{12} = 0 \label{gravity shear eqn3}
\end{align}
The computation in these two channels is very similar to that of the transverse channel and longitudinal channel of the $U(1)$ gauge field in the previous section, so we will be brief.

For the scalar channel, by the substitution $\tilde{\mathbf{h}}_{23}(z) = z^\frac{1}{2}(1-z^2)^{-\frac{1}{2}} w(z)$ we obtain Heun equation in the normal form
\begin{align}
    \big( \partial_z^2 + \frac{\frac{1}{4}-1^2}{z^2} + \frac{\frac{1}{4}-(\frac{m}{2})^2}{(z-1)^2} + \frac{\frac{1}{4}-(\frac{m}{2}i)^2}{(z+1)^2} + \frac{4m^2+p^2+2}{8z(z-1)} + \frac{-4m^2-p^2+2} {8z(z+1)} \big)w(z)=0
\end{align}
with
\begin{align}
    t=-1,a_0=1,a_1=\frac{|m|}{2},a_t=\frac{m}{2}i,a_\infty=0,u=-\frac{p^2+4m^2-2}{8}
\end{align}
With the boundary conditions and the connection relation (\ref{connection relation a0=1}) we find
\begin{align}
    &\tilde{\mathbf{h}}_{23} = \widetilde{\delta\gamma}_{23}\Big[(1+\ldots) +\frac{(p^2+4m^2)^2}{64}\big( 4\psi(1)+5-\sum_{\sigma=\pm,\sigma^{'}=\pm}\psi(-\frac{1}{2}+\sigma\frac{m}{2}+\sigma^{'}a)\nonumber\\
    &+\partial_{a_0}^2 F - \frac{32}{p^2+4m^2}(4a^2-2a^2m^2+\frac{1}{4}m^4 + 4(\partial_tF)^2+(-8a^2+2m^2)\partial_tF-4\partial_tF \partial_t\partial_{a_0}F \nonumber\\
    &+ (-2+4a^2-m^2)\partial_t\partial_{a_0}F) \big) z^2(1+\ldots)\Big]
\end{align}
For the shear channel, plugging (\ref{gravity shear eqn3}) into $\partial_z$(\ref{gravity shear eqn1}), we obtain
\begin{align}
    \big(\partial_z^2-\frac{2z}{1-z^2}\partial_z - \frac{4m^2+p^2(1-z^2)-8z(1-z^2)}{4z(1-z^2)^2}\big) \partial_z \tilde{\mathbf{h}}_{t2} = 0
\end{align}
By the substitution $\partial_z \tilde{\mathbf{h}}_{t2} = (1-z^2)^{-\frac{1}{2}}w(z)$, the equation is transformed to Heun equation of normal form
\begin{align}
     \big( \partial_z^2 + \frac{\frac{1}{4}-(\frac{1}{2})^2}{z^2} + \frac{\frac{1}{4}-(\frac{m}{2})^2}{(z-1)^2} + \frac{\frac{1}{4}-(\frac{m}{2}i)^2}{(z+1)^2} + \frac{4m^2+p^2-10}{8z(z-1)} - \frac{4m^2+p^2+10} {8z(z+1)} \big)w(z)=0
\end{align}
with
\begin{align}
    t=-1,a_0=\frac{1}{2},a_1=\frac{|m|}{2},a_t=\frac{m}{2}i,a_\infty=\frac{3}{2},u=-\frac{p^2+4m^2+10}{8}
\end{align}
With boundary conditions and the connection relation (\ref{connection relation a0=1}) we find
\begin{align}
    &\partial_z \tilde{\mathbf{h}}_{t2}= \frac{2mp\widetilde{\delta\gamma}_{12}-p^2\widetilde{\delta\gamma}_{t2}}{4} (1-z^2)^{-\frac{1}{2}}\big[ w^{(0)}_{\widetilde-} \nonumber\\
    &+ \frac{p^2+4m^2}{4}\big(-2\psi(1)-1+\frac{1}{2}\psi(1+a_1+a)+\frac{1}{2}\psi(1+a_1-a)+\frac{1}{2}\psi(a_1+a)+\frac{1}{2}\psi(a_1-a)\nonumber\\
    &-\frac{1}{2}\partial_{a_0}^2F-\frac{2}{p^2+4m^2}(1+2\partial_t\partial_{a_0}F) \big) w^{(0)}_{\widetilde+}\big]
\end{align}
and furthermore
\begin{align}
    &\tilde{\mathbf{h}}_{t2} = \widetilde{\delta\gamma}_{t2}(1+\ldots) + \frac{p^2+4m^2}{32}(2mp\widetilde{\delta\gamma}_{12}-p^2\widetilde{\delta\gamma}_{t2})\big(-2\psi(1)-1
    +\frac{1}{2}\psi(1+a_1+a)+\frac{1}{2}\psi(1+a_1-a)\nonumber\\
    &+\frac{1}{2}\psi(a_1+a)+\frac{1}{2}\psi(a_1-a)
    -\frac{1}{2}\partial_{a_0}^2F-\frac{2}{p^2+4m^2}(1+2\partial_t\partial_{a_0}F) \big) (z^2+\ldots)\nonumber\\
    &\tilde{\mathbf{h}}_{12} = \widetilde{\delta\gamma}_{12}(1+\ldots) + \frac{p^2+4m^2}{32}(2mp\widetilde{\delta\gamma}_{t2}-4m^2\widetilde{\delta\gamma}_{12})\big(-2\psi(1)-1
    +\frac{1}{2}\psi(1+a_1+a)+\frac{1}{2}\psi(1+a_1-a)\nonumber\\
    &+\frac{1}{2}\psi(a_1+a)+\frac{1}{2}\psi(a_1-a)
    -\frac{1}{2}\partial_{a_0}^2F-\frac{2}{p^2+4m^2}(1+2\partial_t\partial_{a_0}F) \big) (z^2+\ldots)
\end{align}
The coefficient of $z^2$ in the solution of perturbed bulk metric corresponds to the variation of $\mathbf{g}^{(4)}_{ij}$ and hence the variation of the one-point correlator by (\ref{stress tensor 1pt correlator simplified}). Then we can read off the two-point correlators
\begin{align}
    &\langle \tilde{T}_{23}(\omega=\frac{2m}{\rho_0},p) \tilde{T}_{23}(-\omega,-p)\rangle = \frac{1}{2\pi G}\frac{(p^2+\omega^2)^2}{32} \mathcal{C}_3(\omega=\frac{2m}{\rho_0},p) \nonumber\\
    &\langle \tilde{T}_{t2}(\omega=\frac{2m}{\rho_0},p) \tilde{T}_{t2}(-\omega,-p)\rangle = \frac{1}{2\pi G } \frac{p^2+\omega^2}{32}p^2 \mathcal{C}_4(\omega=\frac{2m}{\rho_0},p) \nonumber\\
    &\langle \tilde{T}_{t2}(\omega=\frac{2m}{\rho_0},p) \tilde{T}_{12}(-\omega,-p)\rangle = -\frac{1}{2\pi G} \frac{p^2+\omega^2}{32} \omega p\mathcal{C}_4(\omega=\frac{2m}{\rho_0},p) \nonumber\\
    &\langle \tilde{T}_{12}(\omega=\frac{2m}{\rho_0},p) \tilde{T}_{12}(-\omega,-p)\rangle = \frac{1}{2\pi G} \frac{p^2+\omega^2}{32} \omega^2 \mathcal{C}_4(\omega=\frac{2m}{\rho_0},p)
\end{align}
with
\begin{align}
    &\mathcal{C}_3(\omega=\frac{2m}{\rho_0},p) = \big[ 2\psi(1)+\frac{5}{2} - \frac{1}{2}\sum_{\theta,\sigma=\pm}\psi(-\frac{1}{2}+\theta\frac{m}{2}+\sigma a)\nonumber\\
    &+\frac{1}{2}\partial_{a_0}^2 F - \frac{16}{(\rho_0^2p^2+4m^2)^2}\big(4a^2-2a^2m^2+\frac{1}{4}m^4 + 4(\partial_tF)^2+(-8a^2+2m^2)\partial_tF \nonumber\\
    &-4\partial_tF \partial_t\partial_{a_0}F + (-2+4a^2-m^2)\partial_t\partial_{a_0}F\big) \big]\big|_{t=-1,a_0=1,a_1=\frac{|m|}{2},a_t=\frac{m}{2}i,a_\infty=0,u=-\frac{\rho_0^2p^2+4m^2-2}{8}} \nonumber\\
    &\mathcal{C}_4(\omega=\frac{2m}{\rho_0},p) = \big(2\psi(1)+1
    -\frac{1}{2}\sum_{\theta,\sigma=\pm}\psi(\theta\frac{m}{2}+\sigma a) \nonumber\\
    &+\frac{1}{2}\partial_{a_0}^2F +\frac{2}{\rho_0^2p^2+4m^2}(1+2\partial_t\partial_{a_0}F) \big)\big|_{  t=-1,a_0=\frac{1}{2},a_1=\frac{|m|}{2},a_t=\frac{m}{2}i,a_\infty=\frac{3}{2},u=-\frac{\rho_0^2p^2+4m^2+10}{8}}
\end{align}
The unsolved part for the stress tensor is the sound channel. We have
\begin{align}
    &\partial_z^2 \tilde{\mathbf{h}}_{tt} - \frac{3-5z^2}{2z(1-z^2)}\partial_z \tilde{\mathbf{h}}_{tt} - \frac{1+z^2}{2z} \partial_z(\tilde{\mathbf{h}}_{11}+\tilde{\mathbf{h}}_{22}+\tilde{\mathbf{h}}_{33}) + \frac{-4z+12z^3-p^2(1-z^2)}{4z(1-z^2)^2} \tilde{\mathbf{h}}_{tt} \nonumber\\
    &- \frac{4m^2}{4z(1-z^2)}(\tilde{\mathbf{h}}_{11}+\tilde{\mathbf{h}}_{22}+\tilde{\mathbf{h}}_{33}) + \frac{2mp}{2z(1-z^2)}\tilde{\mathbf{h}}_{t1} = 0 \label{gravity sound eqn 1}\\
    &\partial_z^2 \tilde{\mathbf{h}}_{11} - \frac{3+z^2}{2z(1-z^2)}\partial_z \tilde{\mathbf{h}}_{11} - \frac{1}{2z(1-z^2)}\partial_z \tilde{\mathbf{h}}_{tt} - \frac{1}{2z}\partial_z(\tilde{\mathbf{h}}_{22}+\tilde{\mathbf{h}}_{33}) \nonumber\\
    &-\frac{4m^2}{4z(1-z^2)^2} \tilde{\mathbf{h}}_{11} - \frac{p^2+4z}{4z(1-z^2)^2} \tilde{\mathbf{h}}_{tt} - \frac{p^2}{4z(1-z^2)}(\tilde{\mathbf{h}}_{22}+\tilde{\mathbf{h}}_{33}) + \frac{2mp}{2z(1-z^2)^2} \tilde{\mathbf{h}}_{t1} = 0 \label{gravity sound eqn 2}\\
    &\partial_z^2 (\tilde{\mathbf{h}}_{22}+\tilde{\mathbf{h}}_{33}) - \frac{2}{z(1-z^2)} \partial_z (\tilde{\mathbf{h}}_{22}+\tilde{\mathbf{h}}_{33}) -\frac{1}{z(1-z^2)} \partial_z \tilde{\mathbf{h}}_{tt} - \frac{1}{z}\partial_z \tilde{\mathbf{h}}_{11} \nonumber\\
    &-\frac{4m^2+p^2(1-z^2)}{4z(1-z^2)^2}(\tilde{\mathbf{h}}_{22}+\tilde{\mathbf{h}}_{33}) - \frac{2}{(1-z^2)^2} \tilde{\mathbf{h}}_{tt} = 0 \label{gravity sound eqn 3}\\
    &\partial_z^2 \tilde{\mathbf{h}}_{t1} - \frac{1}{z}\partial_z \tilde{\mathbf{h}}_{t1} - \frac{2mp}{4z(1-z^2)}(\tilde{\mathbf{h}}_{22}+\tilde{\mathbf{h}}_{33}) = 0 \label{gravity sound eqn 4}\\
    &\partial_z^2(\tilde{\mathbf{h}}_{11}+\tilde{\mathbf{h}}_{22}+\tilde{\mathbf{h}}_{33}) + \frac{1}{1-z^2}\partial_z^2 \tilde{\mathbf{h}}_{tt} -\frac{z}{1-z^2} \partial_z(\tilde{\mathbf{h}}_{11}+\tilde{\mathbf{h}}_{22}+\tilde{\mathbf{h}}_{33}) \nonumber\\
    &+ \frac{z}{(1-z^2)^2}\partial_z \tilde{\mathbf{h}}_{tt} + \frac{2}{(1-z^2)^3}\tilde{\mathbf{h}}_{tt} = 0 \label{gravity sound eqn 5}\\
    &2m\partial_z(\tilde{\mathbf{h}}_{11}+\tilde{\mathbf{h}}_{22}+\tilde{\mathbf{h}}_{33}) + \frac{2mz}{1-z^2}\partial_z (\tilde{\mathbf{h}}_{11}+\tilde{\mathbf{h}}_{22}+\tilde{\mathbf{h}}_{33}) -p\partial_z \tilde{\mathbf{h}}_{t1} -\frac{2pz}{1-z^2}\tilde{\mathbf{h}}_{t1} = 0 \label{gravity sound eqn 6}\\
    &p\partial_z(\tilde{\mathbf{h}}_{22}+\tilde{\mathbf{h}}_{33}) + \frac{p}{1-z^2}\partial_z \tilde{\mathbf{h}}_{tt} - \frac{2m}{1-z^2}\partial_z \tilde{\mathbf{h}}_{t1} + \frac{pz}{(1-z^2)^2}\tilde{\mathbf{h}}_{tt} = 0 \label{gravity sound eqn 7}
\end{align}
We don't know how to analytically solve the boundary value problem here. For future reference, we can reduce the sound channel to a five-dimensional first-order equation of variables $\tilde{\mathbf{h}}_{tt},\tilde{\mathbf{h}}_{11}, \frac{\tilde{\mathbf{h}}_{22}+\tilde{\mathbf{h}}_{33}}{2},\tilde{\mathbf{h}}_{t1},\partial_z \tilde{\mathbf{h}}_{t1}$ (a similar equation can be found in \cite{Kovtun:2005ev}), and by the substitution
\begin{align}
    \begin{pmatrix}
    \tilde{\mathbf{h}}_{tt}\\
    \tilde{\mathbf{h}}_{11}\\
    \frac{\tilde{\mathbf{h}}_{22}+\tilde{\mathbf{h}}_{33}}{2}\\
    \tilde{\mathbf{h}}_{t1}\\
    \partial_z \tilde{\mathbf{h}}_{t1}
    \end{pmatrix} = \begin{pmatrix}
        0 & -\frac{1}{3}(1-z^2)^2 & \frac{2}{3}z(1-z^2) & 0 & 0 \\
        -z^2 & 1-z^2 & \frac{2}{3}z & 0 & 0 \\
        \frac{1}{2}z^2 & 0 & -\frac{1}{3}z & 0 & 0 \\
        0 & 0 & 0 & 1-z^2 & 0\\
        0 & 0 & 0 & 0 & z
    \end{pmatrix} H
\end{align}
we can transform the equation into a Fuchsian system of normal form \footnote{The connection relation of local solutions of Fuchsian systems has been studied in the Mathematics literature (for example, see \cite{haraoka2015linear} for very brief introduction), and knowledge of the connection relation in our system will enable us to solve the boundary problem. The main progress we can find in the literature is techniques of ``addition'' and ``middle convolution'' \cite{oshima2011fractional} that can transform a Fuchsian system to a simpler one, and most importantly, if we know the connection relation for one system we know it for the other. We are however not able to simplify our Fuchsian system by these transformations.}
\begin{align}
    \partial_z H = (\frac{M_{0}}{z}+\frac{M_{1}}{z-1}+\frac{M_{-1}}{z+1}) H
\end{align}
with
\begin{align}
    M_0 &= \begin{pmatrix}
        -2 & -\frac{2}{3} & 0 & \frac{p}{3m} & \frac{12m^2+p^2}{6mp} \\
        0 & 0 & 0 & 0 & 0 \\
        0 & -m^2+\frac{p^2}{12} & -1 & mp & 0 \\
        0 & 0 & 0 & 0 & 0\\
        0 & 0 & -\frac{mp}{3} & 0 & 2
    \end{pmatrix}\nonumber\\
    M_1 &= \begin{pmatrix}
        0 & 0 & -\frac{1}{3} & 0 & -\frac{m}{p} \\
        0 &-\frac{1}{2} & 0 & -\frac{p}{2m} & -\frac{p}{4m} \\
        \frac{p^2}{8} & \frac{-1+m^2}{2} & 2 & \frac{p(1-m^2)}{2m} & \frac{p}{4m} \\
        0 & 0 & 0 & -1 & -\frac{1}{2}\\
        -\frac{mp}{4} & 0 & \frac{mp}{6} & 0 & 0
    \end{pmatrix}\nonumber\\
    M_{-1} &= \begin{pmatrix}
        0 & 0 & \frac{1}{3} & 0 & -\frac{m}{p} \\
        0 & -\frac{1}{2} & 0 & -\frac{p}{2m} & -\frac{p}{4m} \\
        \frac{p^2}{8} & \frac{1+m^2}{2} & 2 & -\frac{p(1+m^2)}{2m} & -\frac{p}{4m} \\
        0 & 0 & 0 & -1 & -\frac{1}{2}\\
        \frac{mp}{4} & 0 & \frac{mp}{6} & 0 & 0
    \end{pmatrix}
\end{align}

\section{Summary and discussion}
\label{sec:summary and discussion}
In our study, we calculated holographic Euclidean thermal correlators of the $U(1)$ current and stress tensor for four-dimensional CFTs using the $\mathrm{AdS}_5$ planar black hole, following the approach of GKPW. By utilizing the connection relation of local solutions of the Heun equation, we obtained exact correlators for the $U(1)$ current and stress tensor in the scalar and shear channels.

Extensive research has focused on thermal two-point correlators (thermal spectral functions). Notably, \cite{Kovtun:2005ev} demonstrated the presence of gauge invariants in each channel that diagonalize coupled differential equations. These invariants and their derivatives render the on-shell action quadratic. Thermal two-point correlators have been computed using this formalism numerically or analytically by approximations \cite{Kovtun:2006pf, Teaney:2006nc}. For example in the longitudinal channel the gauge invariant is $E_L = p \tilde{\mathbf{A}}_t - \omega \tilde{\mathbf{A}}_1$ and we have
\begin{align}
    \partial_z^2 E_L - \frac{2\omega^2 z}{(1-z^2)(\omega^2+p^2(1-z^2))} \partial_z E_L - \frac{\omega^2+p^2(1-z^2)}{4z(1-z^2)^2}E_L = 0
\end{align}
This is a Fuchsian differential equation with six singularities. The two singularities $z=\pm\sqrt{1+\frac{\omega^2}{p^2}}$ are apparent singularities since they don't appear in the equation of the fields. One can verify that these apparent singularities cannot be transformed away by a substitution $E_L(z) = P(z)f(z)$ where $P(z)$ is a meromorphic function that does not introduce new singularities. In essence, these apparent singularities remain inherent to the equation. We don't know how to relate this equation to the Heun equation and obtain the exact holographic correlators. From the technical standpoint, we want to work with equations of fields, and in the Euclidean signature, the boundary conditions of fields with gauge/diffeomorphism symmetry are clearly specified. This is the technical reason for our approach of holographic computation, in addition to giving an illustrative example of Euclidean boundary value problems.

Thermal two-point correlators find diverse applications. They encode the information in operator product expansion (OPE) of holographic CFTs. For instance, \cite{Fitzpatrick:2019zqz, Li:2019tpf, Karlsson:2022osn} computed holographic correlators in the OPE limit via near-boundary analysis, extracting OPE coefficients for multi-stress tensors. For integer operator dimension with operator mixing, exact two-point correlators are necessary for complete OPE coefficient extraction. Moreover, if we can analytically continue to the Lorentzian signature, we will get better understanding of the linear response to perturbations in thermal equilibrium, and compute transport coefficients such as shear viscosity, thermal conductivity, and electric conductivity \cite{Policastro:2001yc, Hartnoll:2016apf}, and higher order transport coefficients (see \cite{Kovtun:2018dvd,Grieninger:2021rxd} for formula of second order coefficients in terms of two-point correlators and holographic computation). In addition, we can probe the chaotic dynamics by studying the pole-skipping of the correlators \cite{Grozdanov:2017ajz, Blake:2017ris, Blake:2018leo}. In addition, we consider spherical thermal correlators (scalar case solved in \cite{Dodelson:2022yvn}) and stress tensor correlators from a higher derivative gravity theory as interesting generalizations of our work.

\acknowledgments
We would like to thank Alba Grassi, Cristoforo Iossa, Yun-Ze Li, Hongfei Shu, Ashish Shukla and Yunda Zhang for their helpful discussion. S.H. would appreciate the financial support from the Fundamental Research Funds for the Central Universities and Max Planck Partner Group and the Natural Science Foundation of China (NSFC) Grants No.~12075101 and No.~12235016.

\appendix

\section{Fuchsian ODE, the Heun equation and connection problem} \label{Fuchsian ODE, Heun and connection problem}

In this appendix, we briefly review Fuchsian differential equations, the Heun equation, and its connection relation we used in the computation in the main text.

An ordinary differential equation (ODE) is called Fuchsian if the coefficients are rational functions and all singularities are regular. Eigenvectors of local monodromy constitute a natural basis of local solutions around singularities. When eigenvalues of the local monodromy are all distinct, eigenvectors span the space of local solutions, and they take the form of a series
\begin{align}
    w^{(z_0)}_k = (z-z_0)^{\rho_k} \sum_{i=0}^\infty c_i (z-z_0)^i
\end{align}
where $z_0$ is the singularity, $k$ labels the local solution and the prefactor $(z-z_0)^{\rho_k}$ captures the local monodromy. The characteristic exponents $\rho_k$ are computed as the roots of the indicial equation. We usually adopt the normalization that $c_0=1$. When we have repeated eigenvalues of the local monodromy, that is some characteristic exponents differ by integers, we may need generalized eigenvectors to span the space of local solutions, and they are expressed as series with logarithms. For a second order ODE, we label the two characteristic exponents as $\rho_+,\rho_-$, with $\mathrm{Re}\rho_+ \geq \mathrm{Re}\rho_-$. There is always a series solution without logarithm $w^{(z_0)}_+$ with the exponent $\rho_+$\footnote{One can show that the recursion relation of the coefficients of the series with exponent $\rho_+$ is always non-degenerate.}. If two exponents differ by an integer, the other solution $w^{(z_0)}_-$ to form a basis may contain logarithm. There is also no canonical choice of $w^{(z_0)}_-$ since we can add any constant multiple of $w^{(z_0)}_+$ to $w^{(z_0)}_-$. For computational convenience, we choose the convention that the coefficient of the power $(z-z_0)^{\rho_+}$ is zero in $w^{(z_0)}_-$.

The Heun equation is the second-order Fuchsian ODE with four regular singularities. By M\"obius transformation and substitutions, we can bring it to the normal form
\begin{align} \label{Heun eqn normal form}
    \big(&\partial_z^2 + \frac{\frac{1}{4}-a_0^2}{z^2} + \frac{\frac{1}{4}-a_1^2}{(z-1)^2} + \frac{\frac{1}{4}-a_t^2}{(z-t)^2} - \frac{\frac{1}{2}-a_1^2 - a_t^2 - a_0^2 + a_\infty^2 + u}{z(z-1)} + \frac{u}{z(z-t)}\big) w(z) = 0 
\end{align}
The four singularities with exponents at these points are
\begin{align}
    &z=0, \rho=\frac{1}{2}\pm a_0 \nonumber\\
    &z=1, \rho=\frac{1}{2}\pm a_1 \nonumber\\
    &z=t, \rho=\frac{1}{2}\pm a_t \nonumber\\
    &z=\infty, \rho=-\frac{1}{2}\pm a_\infty
\end{align}
We adopt the convention that $\mathrm{Re}a_0 \geq 0$ etc., so the exponents with plus sign will be the exponent with greater real part $\rho^+$. The connection relation of the local solutions in the generic case (that is, characteristic exponents do not differ by an integer) was studied in \cite{Bonelli:2022ten} by relating the Heun equation to the Belavin-Polyakov-Zamolodchikov (BPZ) equation \cite{Belavin:1984vu} satisfied by conformal blocks with degenerate insertion\footnote{One can also refer to the relevant studies offered by \cite{Cao:2016hvd, He:2017lrg}.} in the Liouville field theory in the semiclassical limit. By the Alday-Gaiotto-Tachikawa (AGT) correspondence, the Liouville correlators can be exactly computed by localization in supersymmetric gauge theories \cite{Alday:2009aq, Nekrasov:2002qd, Nekrasov:2003rj, Alday:2009fs, Nekrasov:2009rc}.
Without losing generality, let $z=0$ and $z=1$ be two adjacent singularities, the connection relation between local solutions around these two points is
\begin{align} \label{w1 w0}
    w^{(1)}_\theta(z) = \sum_{\theta^{'}=\pm} \mathcal{M}_{\theta\theta^{'}}(a_1,a_0;a) e^{(\frac{\theta}{2}\partial_{a_1} - \frac{\theta^{'}}{2}\partial_{a_0})F({a_t \atop a_\infty}a{a_1 \atop a_0};\frac{1}{t})} w^{(0)}_{\theta^{'}}(z)
\end{align}
where
\begin{align}
\mathcal{M}_{\theta\theta^{'}}(a_1,a_0;a) = \frac{\Gamma(-2\theta^{'} a_0)\Gamma(1+2\theta a_1)}{\Gamma(\frac{1}{2} + \theta a_1 - \theta^{'} a_0 + a)\Gamma(\frac{1}{2} + \theta a_1 - \theta^{'} a_0 -a)}
\end{align}
and $F$ is the Nekrasov-Shatashvili function, defined as power series in $\frac{1}{t}$ with combinatorially defined rational functions of other parameters as the coefficients, see Appendix \ref{NS function} for details. The exchange momentum $a$ is to be implicitly determined from the relation
\begin{align}
    u = -\frac{1}{4} - a^2 + a_t^2 + a_0^2 + t\partial_t F
\end{align}
In our computation, the masslessness of the bulk fields leads to a degenerate local monodromy of the Heun equation at $z=0$ (the conformal boundary), that is, two exponents differ by an integer ($a_0$ becomes a half-integer). This degenerate scenario can be derived as a limit of the generic case, as a specific solution to the Heun equation continuously depends on the parameters. The emergence of logarithm and the discontinuity of the local monodromy basis reflect a qualitative change of the local monodromy, rather than a specific solution. The solution $w^{(1)}_+$ remains well-defined and continuously depends on parameters including $a_0$, even when $a_1$ approaches half-integers \footnote{Meanwhile $w^{(1)}_-$ is not continuous when $a_1$ approaches half-integers. When both $a_0$ and $a_1$ are half-integers, the complete connection relation is computed by solving two linear equations obtained from the limits of $w^{(0)}_+$ and $w^{(1)}_+$.}. We proceed to take the limit $a_0 \to \frac{N}{2}, N\in\mathbb{N}$ while keeping other parameters, such as $t,a_1,a_t,a_\infty,a$, fixed \footnote{Another curve in the parameter space can also be chosen to approach the limit, such as fixing $u$, an explicit parameter in the Heun equation, instead of $a$. However, as the connection coefficients explicitly depend on $a$, fixing $a$ yields a relatively simple expression for the limit.}. For $a_0=0$ we have
\begin{align}
    w^{(1)}_{+} = \lim_{a_0\to0}\frac{1}{2a_0} \big[&\frac{\Gamma(1+2a_1)\Gamma(1+2a_0)}{\Gamma(\frac{1}{2}+a_1+a_0+a)\Gamma(\frac{1}{2}+a_1+a_0-a)} e^{(\frac{1}{2}\partial_{a_1} + \frac{1}{2}\partial_{a_0}) F} z^{\frac{1}{2}-a_0}(1+\ldots) \nonumber\\
    &- \frac{\Gamma(1+2a_1)\Gamma(1-2a_0)}{\Gamma(\frac{1}{2}+a_1-a_0+a)\Gamma(\frac{1}{2}+a_1-a_0-a)} e^{(\frac{1}{2}\partial_{a_1} - \frac{1}{2}\partial_{a_0}) F} z^{\frac{1}{2}+a_0}(1+\ldots) \big]
\end{align}
The quantity in the square bracket must vanish when $a_0=0$ for the limit to exist. It indeed vanish because $\partial_{a_0}F|_{a_0=0}=0$ with $F$ being an even function of $a_0$. Then the limit becomes the derivative with respect to $a_0$, and we get
\begin{align} \label{connection relation a0=0}
    w^{(1)}_{+} = &\frac{\Gamma(1+2a_1)}{\Gamma(\frac{1}{2}+a_1+a)\Gamma(\frac{1}{2}+a_1-a)}e^{\frac{1}{2}\partial_{a_1}F} z^\frac{1}{2} \nonumber\\
    &(2\psi(1)-\psi(\frac{1}{2}+a_1+a)-\psi(\frac{1}{2}+a_1-a)+\frac{1}{2}\partial_{a_0}^2 F - \log z + \ldots) \nonumber\\
    =&\frac{\Gamma(1+2a_1)}{\Gamma(\frac{1}{2}+a_1+a)\Gamma(\frac{1}{2}+a_1-a)}e^{\frac{1}{2}\partial_{a_1}F} \nonumber\\
    &\big[ -w^{(0)}_{-} + \big(2\psi(1)-\psi(\frac{1}{2}+a_1+a)-\psi(\frac{1}{2}+a_1-a)+\frac{1}{2}\partial_{a_0}^2 F\big) w^{(0)}_{+} \big]
\end{align}
where $\psi$ denotes the digamma function. For $a_0=\frac{1}{2}$ we find
\begin{align}
    w^{(1)}_{+} = &\lim_{a_0\to\frac{1}{2}} \big[ \frac{\Gamma(1+2a_1)\Gamma(2a_0)}{\Gamma(\frac{1}{2}+a_1+a_0+a)\Gamma(\frac{1}{2}+a_1+a_0-a)} e^{(\frac{1}{2}\partial_{a_1} + \frac{1}{2}\partial_{a_0}) F} \nonumber\\
    &\times z^{\frac{1}{2}-a_0}(1+\frac{-\frac{t}{2}+t(a_0^2+a_1^2+a_t^2-a_\infty^2)+(1-t)u}{(1-2a_0)t}z + \ldots) \nonumber\\
    &+ \frac{\Gamma(1+2a_1)\Gamma(-2a_0)}{\Gamma(\frac{1}{2}+a_1-a_0+a)\Gamma(\frac{1}{2}+a_1-a_0-a)} e^{(\frac{1}{2}\partial_{a_1} - \frac{1}{2}\partial_{a_0}) F} z^{\frac{1}{2}+a_0}(1+\ldots) \big] \nonumber\\
    = &\frac{\Gamma(1+2a_1)}{\Gamma(1+a_1+a)\Gamma(1+a_1-a)}e^{(\frac{1}{2}\partial_{a_1}+\frac{1}{2}\partial_{a_0})F} +\Gamma(1+2a_1)e^{\frac{1}{2}\partial_{a_1}F}z \nonumber\\
    &\times\lim_{a_0\to\frac{1}{2}} \frac{1}{1-2a_0} \big[ \frac{\Gamma(2a_0)}{\Gamma(\frac{1}{2}+a_1+a_0+a)\Gamma(\frac{1}{2}+a_1+a_0-a)} e^{\frac{1}{2}\partial_{a_0} F} \nonumber\\
    &\times\frac{-\frac{t}{2}+t(a_0^2+a_1^2+a_t^2-a_\infty^2)+(1-t)u}{t}z^{\frac{1}{2}-a_0} \nonumber\\
    & - \frac{\Gamma(2-2a_0)}{2a_0\Gamma(\frac{1}{2}+a_1-a_0+a)\Gamma(\frac{1}{2}+a_1-a_0-a)}e^{-\frac{1}{2}\partial_{a_0}F} z^{-\frac{1}{2}+a_0}
    +\ldots \big]
\end{align}
The quantity in the square bracket must vanish when $a_0=\frac{1}{2}$ for the limit to exist, that is, we must have
\begin{align} \label{consistency 2}
    &e^{\partial_{a_0}F}\frac{-\frac{t}{2}+t(a_0^2+a_1^2+a_t^2-a_\infty^2)+(1-t)u}{t}\Big|_{a_0=\frac{1}{2}}\nonumber\\
    &= e^{\partial_{a_0}F}\frac{-\frac{1+t}{4}+ta_0^2+ta_1^2+(1-t)a^2-a_\infty^2-(1-t)t\partial_t F}{t}\Big|_{a_0=\frac{1}{2
}}\nonumber\\
    &=a_1^2 - a^2
\end{align}
By the expansion of $F$
\begin{align}
    F &= \frac{(\frac{1}{4}-a^2-a_t^2+a_\infty^2)(\frac{1}{4}-a^2-a_1^2+a_0^2)}{\frac{1}{2}-2a^2}\frac{1}{t} + O(\frac{1}{t^2})
\end{align}
one can verify (\ref{consistency 2}) holds to the order of expansion. Again, the limit becomes the derivative with respect to $a_0$ and we find
\begin{align} \label{connection relation a0=1/2}
    w^{(1)}_{+} = &\frac{\Gamma(1+2a_1)}{\Gamma(1+a_1+a)\Gamma(1+a_1-a)}e^{(\frac{1}{2}\partial_{a_1}+\frac{1}{2}\partial_{a_0})F} \nonumber\\
    &+\frac{\Gamma(1+2a_1)}{\Gamma(a_1+a)\Gamma(a_1-a)} e^{(\frac{1}{2}\partial_{a_1}-\frac{1}{2}\partial_{a_0})F} z \big[ -2\psi(1)-1+\frac{1}{2}\psi(1+a_1+a)+\frac{1}{2}\psi(1+a_1-a) \nonumber\\
    &+\frac{1}{2}\psi(a_1+a)+\frac{1}{2}\psi(a_1-a)-\frac{1}{2}\partial_{a_0}^2F + \log z - \frac{t+t(1-t)\partial_t\partial_{a_0}F}{2(-\frac{t}{2} + t(a_0^2+a_1^2+a_t^2-a_\infty^2) + (1-t)u)}\big] + \ldots \nonumber\\
    = &\frac{\Gamma(1+2a_1)}{\Gamma(1+a_1+a)\Gamma(1+a_1-a)}e^{(\frac{1}{2}\partial_{a_1}+\frac{1}{2}\partial_{a_0})F} w^{(0)}_{-} \nonumber\\
    &+\frac{\Gamma(1+2a_1)}{\Gamma(a_1+a)\Gamma(a_1-a)} e^{(\frac{1}{2}\partial_{a_1}-\frac{1}{2}\partial_{a_0})F} \big[ -2\psi(1)-1+\frac{1}{2}\psi(1+a_1+a)+\frac{1}{2}\psi(1+a_1-a) \nonumber\\
    &+\frac{1}{2}\psi(a_1+a)+\frac{1}{2}\psi(a_1-a)-\frac{1}{2}\partial_{a_0}^2F  - \frac{t+t(1-t)\partial_t\partial_{a_0}F}{2(-\frac{t}{2} + t(a_0^2+a_1^2+a_t^2-a_\infty^2) + (1-t)u)} \big] w^{(0)}_{+} \nonumber\\
    = &\frac{\Gamma(1+2a_1)}{\Gamma(a_1+a)\Gamma(a_1-a)} e^{(\frac{1}{2}\partial_{a_1}-\frac{1}{2}\partial_{a_0})F} \big[ \frac{t}{-\frac{t}{2} + t(a_0^2+a_1^2+a_t^2-a_\infty^2) + (1-t)u} w^{(0)}_{-} \nonumber\\
    &+\big(-2\psi(1)-1+\frac{1}{2}\psi(1+a_1+a)+\frac{1}{2}\psi(1+a_1-a) +\frac{1}{2}\psi(a_1+a)+\frac{1}{2}\psi(a_1-a) \nonumber\\
    &-\frac{1}{2}\partial_{a_0}^2F 
- \frac{t+t(1-t)\partial_t\partial_{a_0}F}{2(-\frac{t}{2} + t(a_0^2+a_1^2+a_t^2-a_\infty^2) + (1-t)u)} \big) w^{(0)}_{+} \big]
\end{align}
In general, the coefficient $c_N$ in the series solution $z^{\frac{1}{2}-a_0}\sum_{k=0}^\infty c_k z^k$ and the connection coefficient for $w^{(0)}_+$ on the right hand side of (\ref{w1 w0}) simultaneously take $a_0=\frac{N}{2}$ as a pole, so the limit $a_0\to \frac{N}{2}$ always becomes a differentiation with respect to $a_0$. For example,
for $a_0=1$ we have
\begin{align} \label{connection relation a0=1}
    &w^{(1)}_+ = \frac{\Gamma(1+2a_1)}{2\Gamma(-\frac{1}{2}+a_1+a)\Gamma(-\frac{1}{2}+a_1-a)}e^{(\frac{1}{2}\partial_{a_1}-\frac{1}{2}\partial_{a_0})F} \big[
    -\frac{1}{(2-2a_0)c_2|_{a_0=1}} w^{(0)}_{\widetilde-} \nonumber\\
    &+ (2\psi(1)+\frac{5}{2}-\frac{1}{2}\psi(-\frac{1}{2}+a_1+a)-\frac{1}{2}\psi(-\frac{1}{2}+a_1-a)-\frac{1}{2}\psi(\frac{3}{2}+a_1+a)-\frac{1}{2}\psi(\frac{3}{2}+a_1-a) \nonumber\\
    &+\frac{1}{2}\partial_{a_0}^2 F + \frac{\partial_{a_0}((2-2a_0)c_2)}{2(2-2a_0)c_2}|_{a_0=1}) w^{(0)}_{\widetilde+}\big]
\end{align}
We use Mathematica to compute the connection relation in the degenerate case for higher values of $N$.

\section{The Nekrasov-Shatashvili function} \label{NS function}
The Nekrasov-Shatashvili function is defined as
\begin{align}
    F({a_t \atop a_\infty}a{a_1 \atop a_0};\frac{1}{t}) = \lim_{\epsilon\to 0} \epsilon \log \big(&(1-\frac{1}{t})^{-2\epsilon^{-1}(\frac{1}{2}+a_t)(\frac{1}{2}+a_1)} \sum_{\vec{Y}} (\frac{1}{t})^{|\Vec{Y}|} z_v(\vec{a},\vec{Y}) \nonumber\\
    &\prod_{\theta=\pm} z_h(\vec{a},\vec{Y},a_1+\theta a_0)z_h(\vec{a},\vec{Y},a_t+\theta a_\infty)\big)
\end{align}
where $\vec{a} = (a,-a)$ and $\vec{Y}=(Y_1, Y_2)$ is a vector with two Young tableau as the components. $|\vec{Y}|=|Y_1|+|Y_2|$ is the sum of the sizes of the two tableau. $z_h$ and $z_v$ are defined as
\begin{align}
    z_h(\vec{a},\vec{Y},b) =& \prod_{I=1,2} \prod_{(i,j)\in Y_I} (a_I + b + (i-\frac{1}{2}) + \epsilon (j-\frac{1}{2})) \nonumber\\
    z_v(\vec{a},\vec{Y}) =& \prod_{I,J=1,2} \prod_{(i,j)\in Y_I}\prod_{(i^{'},j^{'})\in Y_J} \frac{1}{a_I - a_J - L_{Y_J}((i,j))+\epsilon (A_{Y_I}((i,j))+1)} \nonumber\\
    &\frac{1}{a_I - a_J + (L_{Y_I}((i^{'},j^{'}))+1) - \epsilon A_{Y_J}((i^{'},j^{'}))}
\end{align}
where $L$ stands for the leg length and $A$ stands for the arm length of the site in the tableau.

Our definition of the Nekrasov-
Shatashvili function is essentially the definition in \cite{Dodelson:2022yvn}, with $a_t$ and $a_1$ are swapped and $t$ replaced by $\frac{1}{t}$ in the argument. That's because we consider the connection relation between local solutions around $z=0$ and $z=1$, while in \cite{Dodelson:2022yvn} the pair $z=0$ and $z=t$ was considered.

%\paragraph{Note added.} This is also a good position for notes added after the paper has been written.

% Bibliography

%% [A] Recommended: using JHEP.bst file
\bibliographystyle{JHEP}
\bibliography{reference.bib}

%% or
%% [B] Manual formatting (see below)
%% (i) We suggest to always provide author, title and journal data or doi:
%% in short all the informations that clearly identify a document.
%% (ii) please avoid comments such as "For a review'', "For some examples",
%% "and references therein" or move them in the text. In general, please leave only references in the bibliography and move all
%% accessory text in footnotes.
%% (iii) Also, please have only one work for each \bibitem.

\end{document}